\documentclass[preprint,showpacs,showkeys,preprintnumbers,amsmath,amssymb]{revtex4}

\usepackage{graphicx}
\usepackage{dcolumn}
\usepackage{bm}

\begin{document}


\title{Minimum entropy density method for the time series analysis}

\author{Jeong Won Lee}
\thanks{The first two authors contributed equally to this work.}
\author{Joongwoo Brian Park}
\thanks{The first two authors contributed equally to this work.}
\author{Hang-Hyun Jo}\altaffiliation[Present address: ]{School of Physics,
Korea Institute for Advanced Study, Seoul 130-722, Republic of Korea}
\author{Jae-Suk Yang}\email{yang@kaist.ac.kr}
\altaffiliation[Present address: ]{Department of Physics, Korea
University, Seoul 136-713, Republic of Korea}
\author{Hie-Tae Moon}
\affiliation{Department of Physics, Korea Advanced Institute of
Science and Technology, Daejeon 305-701, Republic of Korea}

\date{\today}

\begin{abstract}

The entropy density is an intuitive and powerful concept to study
the complicated nonlinear processes derived from physical systems.
We develop the minimum entropy density method (MEDM) to detect the
structure scale of a given time series, which is defined as the
scale in which the uncertainty is minimized, hence the pattern is
revealed most. The MEDM is applied to the financial time series of
Standard and Poor's 500 index from February 1983 to April 2006. Then
the temporal behavior of structure scale is obtained and analyzed in
relation to the information delivery time and efficient market
hypothesis.

\end{abstract}


\maketitle

\section{Introduction}

In recent years, physicists have enlarged the research area to many
interdisciplinary fields. Econophysics is one of the active research
areas where many statistical methods are applied to investigate
financial systems. Many analytic methods are introduced, such as the
correlation function, multifractality, minimal spanning tree, and
spin models
\cite{Arthur1997,Mantegna2000,Bouchaud2000,Mandelbrot2001,Kullmann2000,Giada2002}.
The empirical time series in financial markets have also been
investigated by using various methods such as rescaled range (R/S)
analysis to test the presence of correlations \cite{Peters1991} and
detrended fluctuation analysis to detect long-range correlations
embedded in seemingly non-stationary time series
\cite{Peng1994,Liu1999}.

In this paper we focus on how to find a specific time scale in which
a pattern in a time series is revealed most. Since pattern can be
interpreted as the repetitive structure inside the time series we
will be referring to that specific scale as \textit{structure
scale}. To find this structure scale we introduce the minimum
entropy density method, which will be elaborated in detail and
exemplified with the cases of finite periodic time series with
corruption in Section \ref{sect:model}. It is because the periodic
time series is simple and has a repetitive structure among it
definitely. However, our method can be applied to the other time
series as well as the other processes, such as configurations of
spin chain, if they have any certain structures. As an example of
empirical analysis we apply this method to the time series of
S\&P500 index in Section \ref{sect:anal}. The temporal behavior of
the structure scale of the index is obtained and the implications of
the result is analyzed in relation to the information delivery time
and efficient market hypothesis.

\section{Minimum entropy density method\label{sect:model}}
\subsection{Backgrounds\label{sub:back}}

Since our new method for finding the structure scale of a finite
time series is based on the information theory, we start with
briefly explaining the concepts in the information theory according
to Ref. \cite{Feldman1998}. Firstly, we consider a process given by
an infinitely consecutive discrete random variables,
$\overleftrightarrow{X}=\cdots X_{-1}X_0X_1X_2\cdots$, where each
$X_i$ may take the value $x_i$ drawn from a finite countable set $A$
of size $k$. The probability distribution of a block of $L$
consecutive random variables $X^L=X_i,\cdots,X_{i+L-1}$ is taken as
the set of joint probabilities of $L$ consecutive values
$\Pr(x^L)=\Pr(x_i,\cdots,x_{i+L-1})$ for all $k^L$ possibilities.
Then the Shannon entropy for the above $L$-block variable $X^L$ is
defined as
\begin{equation}
H(L)=- \sum_{x_1\in A} \cdots \sum_{x_L\in A}
\Pr(x_1,\cdots,x_L)\log_2 \Pr(x_1,\cdots,x_L), \label{eq:HL}
\end{equation}
which measures the uncertainty or randomness in the process. $H(L)$
is a monotonically increasing function of $L$ because the more
relevant information can be extracted from the time series for the
larger $L$. We can measure the entropy of the infinite process
$\overleftrightarrow{X}$ by taking $L\rightarrow\infty$. However,
$H(L)$ may diverge as $L$ goes to infinity, so an entropy density is
introduced as follows:
\begin{equation}
h_\mu\equiv \lim_{L\rightarrow \infty}\frac{H(L)}{L},
\label{eq:hmu0}
\end{equation}
equivalently
\begin{equation}
h_\mu = \lim_{L\rightarrow \infty}\{H(L+1)-H(L)\}. \label{eq:hmu}
\end{equation}

If the process $\overleftrightarrow{X}$ contains a periodic
structure, for a sufficiently large $L$ (larger than the period)
increasing $L$ does not give us any more information. In this case
the entropy density becomes $0$. On the other hand, if the process
has been generated totally randomly, $\Pr(x^L)=k^{-L}$ for all $k^L$
possibilities, then $H(L)=L\log_2k$ and consequently
$h_\mu=\log_2k$, which is the maximum value of the entropy density.
Therefore the repetitive structure embedded in the process makes the
entropy density lower than that of a more random process. In
addition the entropy density can be interpreted as the uncertainty
of a given variable when all the preceding variables are known. If
there exists a repetitive structure in the process, the knowledge of
all the previous information will greatly decrease the uncertainty
of the next variable.

Since the finite size of the empirical data sets directly a limit to
the block size $L$, we need the finite-$L$ approximation to the
thermodynamic entropy density $h_\mu$ as follows:
\begin{equation}
h_\mu(L) \equiv H(L)-H(L-1),\quad L=1,2,\cdots, \label{eq:hmuL}
\end{equation}
where $H(0)$ is set to $0$. Actually all the processes we deal with
through this paper are finite, hence only $h_\mu(L)$ matters other
than $h_\mu$. By the way, unless $L$ is large enough to fully detect
the structure in the process, $h_\mu(L)$ would overestimate the
randomness of the process. Therefore, as $L$ increases $h_\mu(L)$
converges to $h_\mu$.

\subsection{Method\label{sub:method}}

For the entropy density analysis we coarse-grain the data with an
appropriate scale and then digitize the continuous amplitudes into
discrete values. Let us consider a temporal data set $Y(t)$ as a
function of discrete time steps $t=0,\cdots,S-1$. Once the scale $s$
to grain the data is given, then the resulting time series has
$N=S/s$ equally spaced measurements. For the digitization we set a
countable set $A$ to the smallest and simplest set of size $k=2$,
such as $\{0,1\}$, among the various alternatives. In other words
the original data set $Y(t)$ changes into the binary time series
$F_n$ by the following process:
\begin{equation}
F_n \equiv \theta \left(Y(sn+s)-Y(sn)\right), \quad 0\leq n\leq
N-2,\label{eq:Fn}
\end{equation}
where $\theta(x)$ is a Heaviside step function. $F_n$ gets the value
of $0$ if the value of measurement has decreased after the interval
$s$ and does the value of $1$ otherwise. To make clear the effect of
choosing $s$ on the coarse-grained data set, consequently on the
entropy density we define $h_\mu(s,L)$ as the entropy density of the
process coarse-grained with scale $s$, which plays a key role in our
method.

The minimum entropy density method (MEDM) is based on the assumption
that the pattern in the time series is revealed most when the time
series is coarse-grained with the structure scale defined as the
scale minimizing the entropy density. Most empirical time series for
the complex systems are usually contaminated by the high frequency
noise and we want to get the noiseless signal or intrinsic structure
from the scratches. Provided that a time series can be described by
a characteristic time scale $s_c$, if the smaller scale than $s_c$
is used for the analysis the time series looks more random due to
the high frequency noise. On the other hand if the larger scale than
$s_c$ is used we overlook the time series so that we fail to get the
structure, hence the time series looks more random too. In short, by
finding the scale minimizing the entropy density we can get the
characteristic scale $s_c$.

For the first step of MEDM we decide the range of coarse-graining
scale, usually set to $[1,s_{max}]$. Then, before finding the
structure scale $s^*$ minimizing $h_\mu(s,L)$ by tuning $s$ we
should determine the appropriate value of $L$. The first of two
criteria for choosing $L$ sets the upper bound of $L$:
\begin{equation}
L<\log_k N,\label{eq:logkN}
\end{equation}
where $N$ is the number of data points $S/s$ and $k$ is the size of
the set $A$. As $L$ increases, the more relevant information can be
extracted from the process while the average number of realizations
for each possibility of $L$-block variable decreases fast as $N/k^L$
for finite $N$. Therefore, for the significant analysis $L$ should
be limited by a condition that the average number of realizations
for each possibility of $L$-block variable should be at least one:
$N/k^L>1$, equivalent to Eq. (\ref{eq:logkN}). For the
mathematically rigorous arguments see Refs.
\cite{Shalizi2004,Marton1994}.

The second criterion is to determine the convergence range of $L$ in
which $h_\mu(s,L)$ for some $s$ converges to $h_\mu$. However,
without the knowledge of $h_\mu$ it is not clear to see whether
$h_\mu(s,L)$ converges to $h_\mu$ or not. If the value of $s^*$ does
not depend on $L$ we can determine the structure scale $s^*$ even
though $h_\mu(s,L)$ does not converge yet. On the other hand, if the
value of $s^*$ varies according to $L$ in general we have to find
the convergence range of $L$, which can be practically defined as
the range where the landscape of $h_\mu(s,L)$ is approximately flat.
Once such a convergence range exists for any $s$, it would be enough
to determine $s^*$ for that range of $L$ because the aim of this
paper is to compare the entropy densities for different scales for a
fixed value of $L$, not to get the better approximation of entropy
densities.

The MEDM that we have discussed so far can be summarized into four
main steps including how to choose an appropriate value of $L$:
\begin{enumerate}
\item
Decide the range of coarse-graining scale, usually set to
$[1,s_{max}]$.
\item
For each $s$ in that range, transform the given time series $Y(t)$
into $k$-ary time series, for example, by Eq. (\ref{eq:Fn}).
\item
Choose the appropriate value of $L^*$:
\begin{enumerate}
\item
$L$ should be lower than $\log_kN$, where $N=S/s_{max}$.
\item
$L$ should be chosen inside the convergence range where the
landscape of $h_\mu(s,L)$ is flat for the whole range of $s$.
\end{enumerate}
\item
Find the structure scale $s^*$ minimizing the entropy density
$h_\mu(s,L^*)$ by tuning $s$.
\end{enumerate}
For the last step of MEDM there may be more than one minimum in the
landscape of $h_\mu(s,L^*)$, which will be discussed with the
examples in the next Subsection.

\subsection{Model examples\label{sub:example}}

\begin{figure}[!ht]
\centerline{\includegraphics[scale=1]{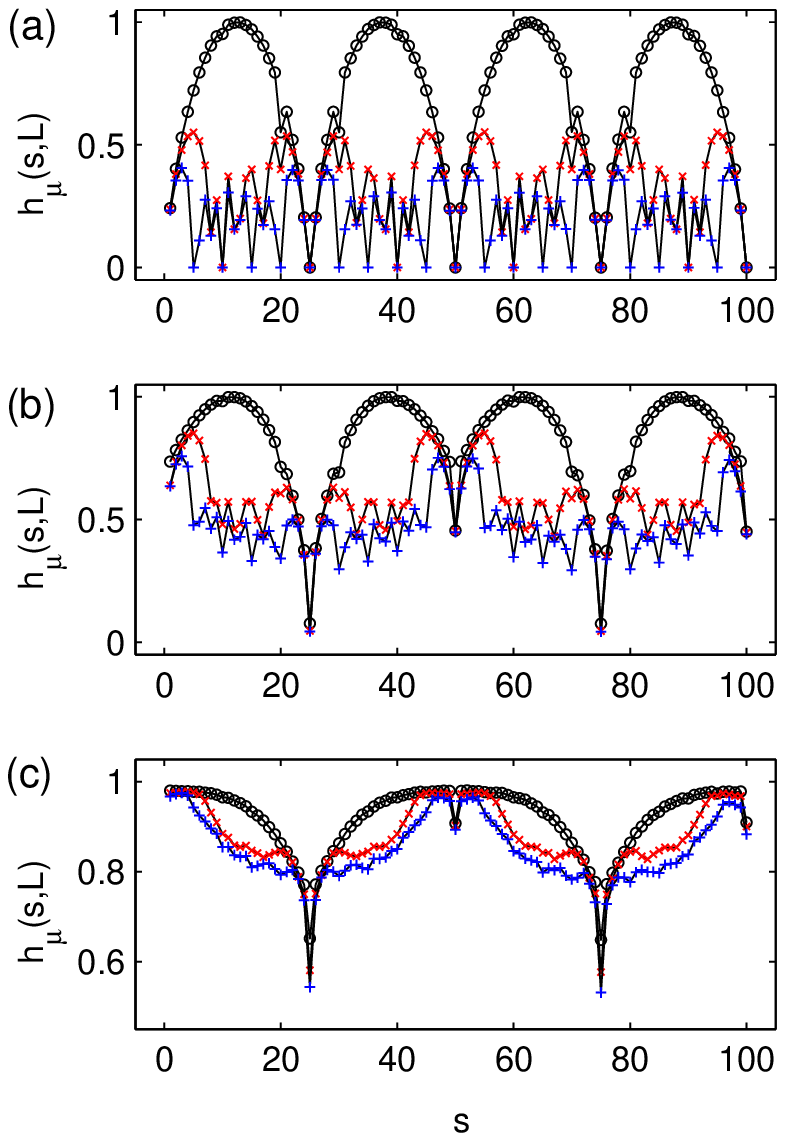}}
\caption{(Color online) The landscapes of entropy densities
$h_{\mu}(s,L)$ of the periodic time series as functions of scale $s$
with block size $L=2$ (black circles), $4$ (red crosses), and $6$
(blue plus signs), respectively. The fraction of corrupted data
points $r$ is $0$ (a), $0.1$ (b), and $0.5$ (c), respectively. For
(b) and (c) each point is averaged over $50$ realizations and for a
clear view we plotted $s$ to $100$ not to $s_{max}=200$.}
\label{fig1}
\end{figure}

The MEDM is applied to the finite periodic time series with
corruption because the periodic time series is simple and has a
repetitive structure among it definitely. We consider the following
series: for each time step $t$,
\begin{equation}
Y(t) = \left\{
\begin{array}{ll}
\cos^2\left(\frac{\pi}{p}t\right) & \quad \textrm{with probability} \, 1-r\\
\eta & \quad \textrm{with probability} \, r
\end{array}
\right., \label{eq:Ytcorr}
\end{equation}
where $p$ is the period of $Y(t)$, $\eta$ is a random number
uniformly drawn from $[0,1]$, and $r$ represents the fraction of
corrupted data points. We set the data size $S$ to $10^5$, $p$ to
$50$, $k$ to $2$, and $s_{max}$ to $200$, \textit{i.e.} the $200$
different sets of binary time series are constructed by Eq.
(\ref{eq:Fn}). Then for a few cases of $r$ the entropy densities for
the whole range of $s$ and for $L\leq 8< \log_2 \frac{10^5}{200}$
are calculated, as partly shown in Fig. \ref{fig1}.

\begin{figure}[!ht]
\centerline{\includegraphics[scale=1]{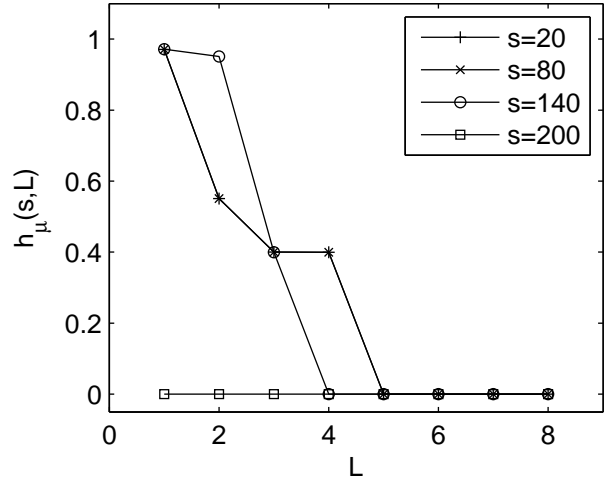}} \caption{The
entropy densities $h_{\mu}(s,L)$ of the time series coarse-grained
with scales $20$ (plus signs), $80$ (crosses), $140$ (circles), and
$200$ (squares) when the fraction of corrupted data points is $0.1$.
Each point is averaged over $50$ realizations. There exists a
convergence range of $L$ in $[5,8]$. } \label{fig2}
\end{figure}

When there is no corrupted data, \textit{i.e.} $r=0$, for $L\leq 3$
the global minima of $h_\mu(s,L)$ turn out to be $0$ for the
multiples of $s=25=p/2$. As $L$ increases there appears the
additional global minima for the other values of $s$. Finally the
entropy densities $h_\mu(s,L)$ for the multiples of $s=5$ become the
global minima when $L=5$ and even when $L>5$. This implies that
$L\geq 5$ is the convergence range of $L$ as shown in Fig.
\ref{fig2}. One can criticize that for the case of $s=1$, where the
binary time series becomes $F=0^{25}1^{25}0^{25}1^{25}\cdots$, the
pattern $0^{25}1^{25}$ can be completely revealed by taking $L$
larger than $50$. But it is contradictory to the first criterion in
Eq. (\ref{eq:logkN}) when given the finite time series. Instead of
taking $L$ as large as possible, we can get more relevant results by
tuning the scale $s$ even for the small values of $L$ guaranteeing
the significance of analysis.

If the corruption is taken into account, \textit{i.e.} in cases of
$r=0.1$ and $0.5$, the local minima for the odd multiples of $s=p/2$
become distinctive among other minima (Fig. \ref{fig1} (b) and (c)).
Moreover, the values of distinctive local minima in the landscape of
$h_\mu(s,L)$ turn out to be independent of $L$ so that the structure
scales $s^*$ are successfully determined and hence it is not
necessary to specify the convergence range of $L$ as well as $L^*$.

Then why do the entropy densities for the odd multiples of $s=p/2$
remain minimized? The corruption definitely destroys the periodicity
of time series and increases the randomness, therefore the overall
values of local minima of $h_\mu(s,L)$ get larger than those for the
case of $r=0$. However, the effect of corruption is not uniform. At
first, without corruption $Y(sn)$ in Eq. (\ref{eq:Fn}) for the odd
multiples of $s=p/2$ take the extreme values of $Y(t)$, precisely
$Y(pn/2)=1$ for $n$ even and $Y(pn/2)=0$ for $n$ odd. Therefore the
flip probability, defined as the probability that the sign of
argument $Y(sn+s)-Y(sn)$ in Eq. (\ref{eq:Fn}) is flipped due to the
corruption, is the least among for the other values of $s$. For
example, if $Y(pn/2)$ remains unchanged as either $0$ or $1$ while
$Y(pn/2+p/2)$ is replaced by a random number in $[0,1]$, the flip
probability is $0$. On the other hand, for the case of $s=p$, if
$Y(pn)$ remains unchanged as $1$ while $Y(pn+p)$ is replaced by a
random number in $[0,1]$, the flip probability is $1$. As a result
one can expect that when the periodic function is corrupted by
noise, the most robust scale is not the period $p$ and its multiples
but $p/2$ and its odd multiples. Based on this argument one can say
that the existence of more than one distinctive local minimum
naturally comes from the repetitive structure of the original
function $Y(t)$, and that the patterns can appear in the different
scales simultaneously.

\begin{figure}[!ht]
\centerline{\includegraphics[scale=1]{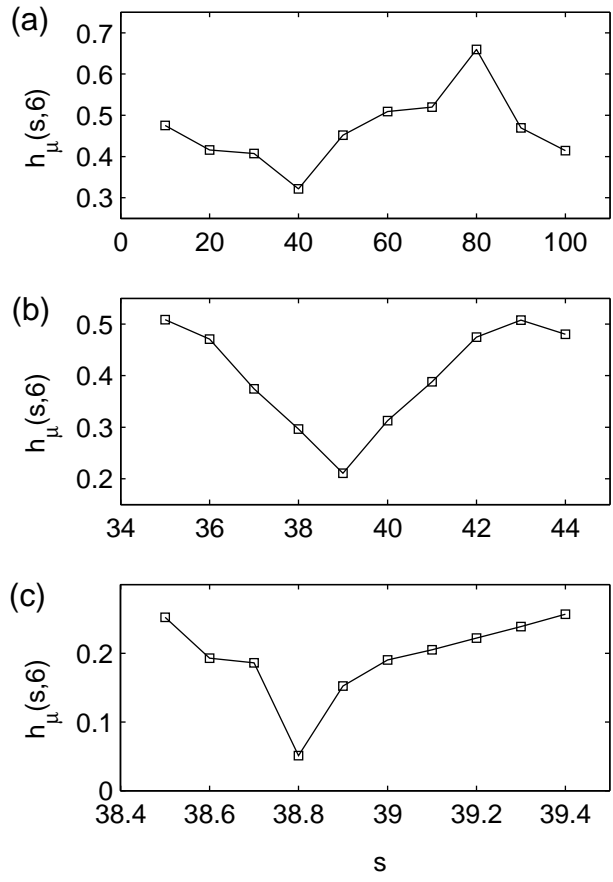}} \caption{The
entropy densities $h_\mu(s,L=6)$ measured in the precision of $10$
(a), $1$ (b), and $0.1$ (c), respectively.} \label{fig3}
\end{figure}

For more general application to a continuous time series we can take
the finer scales to increase the precision of measuring the
structure scale. To show an efficient way to fine-tune $s$ we
consider a corrupted periodic function with non-integer period, for
example, for a continuous time $t$,
\begin{equation}
Y(t) = \left\{
\begin{array}{ll}
\cos^2\left(\frac{\pi}{77.6}t\right) & \quad \textrm{with probability} \, 0.9\\
\eta & \quad \textrm{with probability} \, 0.1
\end{array}
\right.. \label{eq:Ytcorr776}
\end{equation}
To measure the structure scales (the odd multiples of $s=p/2$ for
the case of discrete periodic functions), $s$ should be smaller than
$0.1$. Instead of scanning the whole range of $s$, such as from
$0.1$ to $100.0$, by the increment of $0.1$ we tune $s$ in a larger
scale first and then move down to the smaller scales. The value of
$L$ is set to $6$ according to the MEDM. Figure \ref{fig3}(a) shows
the entropy densities for various $s$ in the order of $10$. The
minimum of the entropy density occurs at $s=40$. We narrow the
variation of $s$ down to $1$ around $40$. Then the minimum of the
entropy density occurs at $s=39$ in Fig. \ref{fig3}(b) and we repeat
the same process again. Finally, in Fig. \ref{fig3}(c) we obtain
$s^*=38.8$ minimizing the entropy density, which is exactly a half
period ($p/2$).

\begin{figure}[!ht]
\centerline{\includegraphics[scale=1]{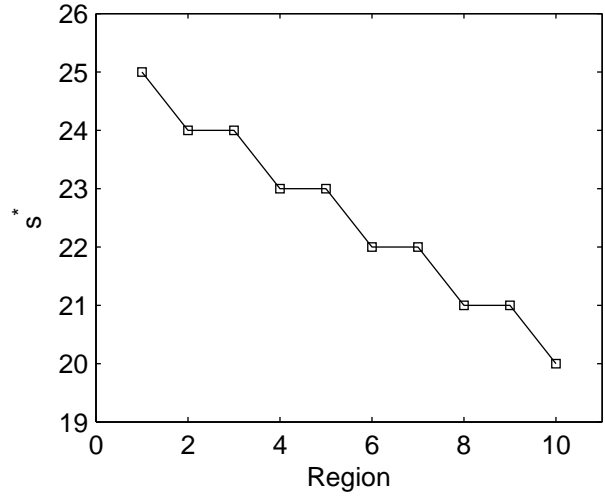}} \caption{The
temporal behavior of the structure scale $s^*$, where each point
represents the $s^*$ for each partitioned region.} \label{fig4}
\end{figure}

Finally, the MEDM can be applied to a periodic function with varying
period by dividing the given time series into several regions and
applying MEDM to each of them. Here we consider a periodic function
with linearly decreasing period: for a continuous time $t$,
\begin{eqnarray}
Y(t) &= &\left\{
\begin{array}{ll}
\cos^2\left(\frac{\pi}{p(t)}t\right) & \quad \textrm{with probability} \, 0.9\\
\eta & \quad \textrm{with probability} \, 0.1
\end{array}
\right.,\\
p(t)&=&p_1+\frac{p_2-p_1}{S}t,\label{eq:Ytcorrpt}
\end{eqnarray}
where the period continuously decreases from $p_1$ to $p_2$. We set
$p_1$ to $50$, $p_2$ to $40$, and $S$ to $10^5$, respectively. The
total time series is divided into $10$ regions and the MEDM is
applied to each of them. For all the regions we tune $s$ in an order
of $1$ and set $L$ to $6$ after testing in a way we described
before. Figure \ref{fig4} shows that the smallest structure scale
$s^*$ decreases from $25=p_1/2$ in the first region to $20=p_2/2$ in
the last one. These $s^*$s are exactly the half periods of the
starting and ending parts of the original function. If we divide the
time series into more regions and use the finer scales, then the
resultant temporal behavior of structure scale gets closer to
$p(t)/2$, where $p(t)$ is defined in Eq. (\ref{eq:Ytcorrpt}), than
before.


\section{Empirical data analysis\label{sect:anal}}

Now we apply the MEDM to analyze the financial time series of the
S\&P500 index from year 1983 to 2006. We used the tick-by-tick data.
It is reasonable to think that the structure scale of S\&P500 index
for $24$ years would change from time to time. Hence the formalism
of the last example in the previous Section is used. It should be
noted that although the time series of the S\&P500 index is not
periodic, we can always measure the structure scale using MEDM
whenever the series has patterns.

The total time span of the index data from February 1983 to April
2006 is divided into $279$ regions, \textit{i.e.} each region for
each month. For each region the structure scale is obtained then the
temporal behavior of it will be analyzed. The unit of
coarse-graining scale $s$ is set to $1$ tick, the finest resolution
of the empirical S\&P500 index data. On average there are $4$ ticks
in one minute though the real time intervals between adjacent ticks
are not equally distributed. One reasonable way to fix this problem
is to obtain the structure scale $s^*_{tick}$ in a unit of tick for
each month and multiply it by the average real time interval $\bar
\tau_{tick}$ between ticks within that month. The resulting value
$s^*=s^*_{tick}\cdot \bar \tau_{tick}$ will be the structure scale
in a unit of time for each month.

\begin{figure}[!ht]
\centerline{\includegraphics[scale=1]{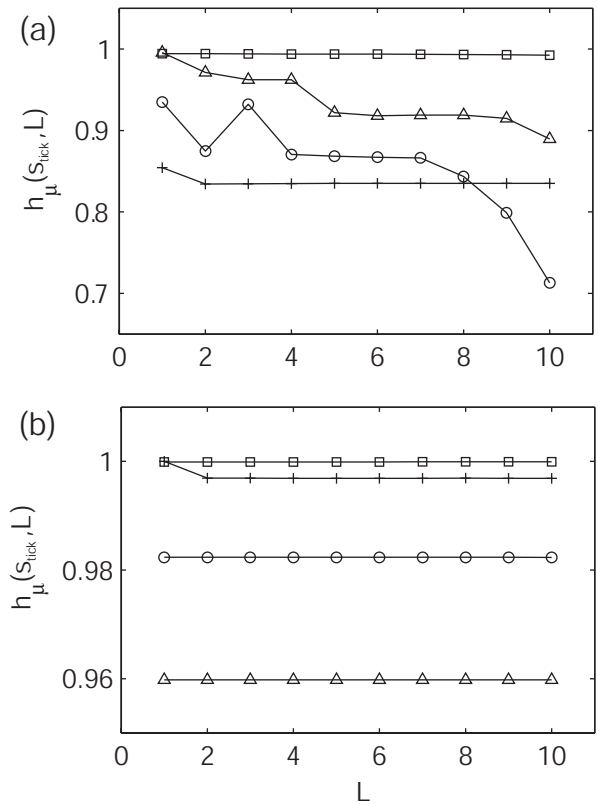}} \caption{ The
entropy densities $h_\mu(s,L)$ of S\&P500 index in February 1983 (a)
and in April 2006 (b) using the time series with scale $s_{tick}=1$
(squares), $2$ (triangles), $5$ (circles), and $10$ (plus signs),
respectively.} \label{fig5}
\end{figure}

\begin{figure}[!ht]
\centerline{\includegraphics[scale=1]{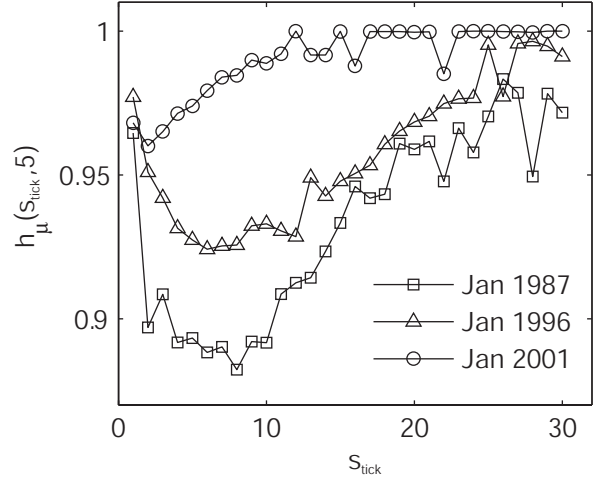}} \caption{ The
entropy densities $h_\mu(s,L=5)$ of S\&P500 index measured in
January 1987 (squares), January 1996 (triangles), and January 2001
(circles), respectively.} \label{fig6}
\end{figure}

\begin{figure}[!ht]
\centerline{\includegraphics[scale=1]{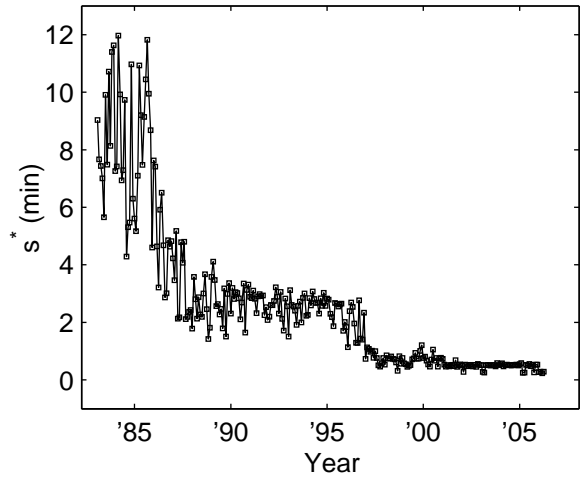}} \caption{The
temporal behavior of the structure scale $s^*$ of the S\&P500 index
measured monthly in a unit of time.} \label{fig7}
\end{figure}

Then we follow the four main steps of the MEDM to measure the
structure scale of tick every month. For the first step the range of
$s_{tick}$ is set to $1$ tick to $30$ ticks. The tick series with
$s_{tick}=30$ has less than $900$ data points each month. By Eq.
(\ref{eq:logkN}) the upper bound of $L$ is $9$. Considering the
second criterion of choosing $L$, we set $L^*$ to $5$ by finding the
convergence region of $L$ for the whole range of $s_{tick}$. Figure
\ref{fig5} shows the landscapes of entropy densities
$h_\mu(s_{tick},L)$ only for the regions of February 1983 and April
2006. For the third step the structure scale $s^*_{tick}$ minimizing
$h_\mu(s_{tick},5)$ is determined for each month. Three examples are
shown in Fig. \ref{fig6}, where $h_\mu(s_{tick},5)$ is minimized at
$s_{tick}=8$ for January 1987, at $s_{tick}=6$ for January 1996, and
at $s_{tick}=2$ for January 2001, respectively. Unlike the case with
the periodic time series, for each region there is only one
structure scale over the range of $s$. After finding all the
$s^*_{tick}$ we convert them into the real time scales by
multiplying the average time interval between ticks for each month.
Finally we get the temporal behavior of the structure length $s^*$
as shown in Fig. \ref{fig7}. During 1980's and 1990's $s^*$
decreases slowly but declines fast after late 1990's.

We analyze the meaning of this result by considering the time scale
by which the information flows among interacting agents in the stock
market. The stock market price changes only when the agents in the
stock market buy or sell. Since the agents make decisions based on
the information they get, the information delivery time can be one
of the most important factors for the changing rate of price. The
information delivery time (IDT), defined as the time taken for the
delivery of information from sources to agents, is assumed to be
proportional to the average price change cycle. If the entropy
density of the time series is measured with scale $s$ smaller than
the IDT, it would be relatively high because the coarse-grained time
series looks more random due to the high frequency noise. On the
other hand, if $s$ is larger than IDT, we overlook the pattern
embedded in the time series so fail to detect the structure scale
and the coarse-grained time series looks more random too. Therefore,
if the optimally closest scale to the IDT is used to detect the
patterns in the time series, the entropy density for that scale
would be minimized due to the repetitive structure of the price
change. Consequently,
\begin{equation}
IDT \approx s^*.\label{eq:IDT}
\end{equation}

The long-term decrease of $s^*$ from year 1986 to 2006 in Fig.
\ref{fig7} can be interpreted as the decrease of the information
delivery time. The value of $s^*$ suddenly jumps down around year
1997, when the Internet was starting to spread widely, the fraction
of online traders increased exponentially. These influenced the IDT
of the stock market to become much shorter.

Since there does not exist any standardized way to measure the
information delivery time, we suggest $s^*$ as one of candidates to
measure it. IDT can be also used to measure the efficiency of the
stock market: if the market is idealized with efficient market
hypothesis (EMH) \cite{Mantegna2000}, then IDT will become $0$. In
addition from our quantitative analysis IDT of the S\&P500 index is
about $17$ seconds in year 2006.

\section{Conclusions}

In this paper we have developed the minimum entropy density method
(MEDM) to detect the structure scale of a given time series. This
method is based on the assumption that the pattern in the time
series is revealed most when the time series is coarse-grained with
the structure scale defined as the scale minimizing the entropy
density. We also showed that the MEDM is useful to detect the
repetitive structures in the various time series if they have
certain patterns.

Additionally, by applying the MEDM to the financial time series of
S\&P500 index we identified that the time scale with the most
patterns showing, has decreased for the last twenty years. In other
words the information flows faster than before. The MEDM has also
been applied to Korea Composite Stock Price Index (KOSPI) from April
1992 to June 2003 with $1$ minute time interval \cite{Lee2006}. The
structure scale of the KOSPI index, which can be interpreted as the
IDT, had also decreased for ten years similar to S\&P500 index. We
believe this effect is real, considering that the Internet trading
has become popular recently, which we think is one of the main
factors of decreasing the IDT, in both U.S. and Korean stock market.
Also, Yang and colleagues \cite{Yang2006} used the microscopic spin
model to investigate the financial market and identified that the
change of log-return distributions of financial stock markets can
result from the increasing velocity of information flow, which
implies that the IDT becomes shorter than before. Since IDT measures
the efficiency of the stock market, by quantitative analysis we
conclude that the efficiency of the U.S. stock market dynamics
became close to EMH.






\newpage

\end{document}